\documentclass[useAMS,usenatbib]{mn2e}

\usepackage{times}
\usepackage{rotating}
\usepackage{amsmath}
\usepackage{amssymb}
\usepackage[dvipsnames]{xcolor}
\usepackage{graphicx}
\usepackage{graphics}
\usepackage{upgreek}
\usepackage{hyperref}
\hypersetup{
     colorlinks   = true,
     citecolor    = blue
}

\def\apj{{ ApJ}}
\def\apjl{{ApJL}}

\def\aap{{ A\&A}}

\def\mnras{{ MNRAS}}
\def\araa{{ ARA\&A}}

\def\nat {{ Nature}}

\def\physrep{{ Physics Reports}}

\def\prd{{ Phys. Rev. D}}

\def\pasa{ PASA}

\def\mr{\mathrm}
\def\pi{\uppi}

\def\d{\mathrm{d}}
\def\para{\parallel}
\def\b{\boldsymbol}

\def\me{m_{\rm e}}

\def\bphi{\beta_{\phi}}

\makeatletter
\newcommand{\lambdabar}{{\mathchoice
  {\smash@bar\textfont\displaystyle{0.25}{1.2}\lambda}
  {\smash@bar\textfont\textstyle{0.25}{1.2}\lambda}
  {\smash@bar\scriptfont\scriptstyle{0.25}{1.2}\lambda}
  {\smash@bar\scriptscriptfont\scriptscriptstyle{0.25}{1.2}\lambda}
}}
\newcommand{\smash@bar}[4]{%
  \smash{\rlap{\raisebox{-#3\fontdimen5#10}{$\m@th#2\mkern#4mu\mathchar'26$}}}%
}

\long\def\symbolfootnote[#1]#2{\begingroup%
\def\thefootnote{\fnsymbol{footnote}}\footnote[#1]{#2}\endgroup}
\newcommand{\gae}{\lower 2pt \hbox{$\, \buildrel {\scriptstyle >}\over {\scriptstyle
\sim}\,$}}
\newcommand{\lae}{\lower 2pt \hbox{$\, \buildrel {\scriptstyle <}\over {\scriptstyle
\sim}\,$}}

\makeatother

\newcommand{\myemail}{wenbinlu@caltech.edu}

\title[Maximum Luminosity of FRBs]{The Maximum Luminosity of
Fast Radio Bursts}
\author[Lu \& Kumar]
  {Wenbin Lu$^{1,2}$\thanks{\myemail} and Pawan
    Kumar$^2$\thanks{pk@astro.as.utexas.edu}\\ 
$^1$TAPIR, Walter Burke Institute for Theoretical Physics, Mail Code
  350-17, Caltech, Pasadena, CA 91125, USA\\
  $^2$Department of Astronomy, University of Texas at Austin, Austin,
TX 78712, USA}

\begin{document}
\label{firstpage}
\maketitle

\begin{abstract}
Under the assumption that fast radio bursts (FRBs) are from
coherent curvature emission powered by the
dissipation of magnetic energy in the magnetosphere of neutron
stars, we show that there is a maximum isotropic equivalent luminosity
$L_{\rm max}\sim (2\times 10^{47}\,
\mr{erg\,s^{-1}})\,\mr{min}(\rho_6^2, B_{16}\rho_6^{4/3}\nu_9^{-2/3})$, where
$\rho_6=\rho/10\,$km is the curvature radius of the magnetic field
lines near the source region, $B_{16} = B/10^{16}\,$G is the local magnetic
field strength, and $\nu_9 = \nu/$GHz is the FRB wave frequency. This is
because the electric field responsible for accelerating the emitting
particles becomes close to the quantum critical strength and
is then quickly shielded by Schwinger pairs within a
nano-second. Future observations should be able to measure this
cut-off luminosity and hence provide a unique probe of the source
location and magnetic field strength. We discuss the effects of  
$L_{\rm max}$ on the observed flux distributions for repeating bursts
from the same object and for the entire FRB population.

\end{abstract}


\section{Introduction}
A major breakthrough in understanding the nature of
fast radio bursts \citep[FRBs,][]{2007Sci...318..777L, 2013Sci...341...53T}
came when the repeater FRB121102 was precisely
localized to be in a dwarf galaxy at redshift $z=0.193$
\citep{2014ApJ...790..101S, 2016Natur.531..202S, 2017Natur.541...58C,
  2017ApJ...834L...7T, 2017ApJ...834L...8M, 
  2017ApJ...843L...8B}. Confirmation of the cosmological origin of FRBs
means that they are highly energetic events seen in the radio
band. The bursts from the repeater show a power-law distribution of
isotropic equivalent 
luminosities $\d N/\d L\propto L^{-\beta}$ in the range from
$\sim$$10^{40}$ to 
$\sim$$10^{43}\rm\,erg\,s^{-1}$ and $\beta\sim 1.7$ \citep{2016ApJ...833..177S,
  2017ApJ...850...76L, 2017MNRAS.472.2800H, 2017ATel10693....1O}. The
luminosity distribution of other so-far non-repeating FRBs is less certain due
to poor localization and unknown distances. If the
Milky-Way-subtracted dispersion measures (DMs) are dominated by 
the intergalactic medium, their isotropic equivalent luminosities range from
$\sim$$10^{42.5}$ to $\sim$$10^{44.5}\rm\,erg\,s^{-1}$ \citep[see the FRB catalog
by][]{2016PASA...33...45P}, with FRB160102 \citep{2018MNRAS.475.1427B}
and FRB170107 \citep{2017ApJ...841L..12B} being the brightest ones
detected so far. We note that these luminosities may not correspond to
intrinsic values because (i) the reported peak fluxes in most cases
are based on the assumption that the burst occurred at the beam center,
 (ii) many FRBs are temporarily broadened due to multi-path propagation
\citep{2017arXiv171008026R}, and (iii) lensing by plasma structures in the
host galaxy could further introduce magnification biases
\citep{2017ApJ...842...35C}.

Many models have been proposed to explain FRBs based on considerations
of their event rate, duration and energetics. They generally fall into two
categories \citep[see][for recent reviews of these
models]{2016MPLA...3130013K, 2018arXiv180409092K}: emission from a
relativistic outflow which dissipates its 
energy at large distances from the central compact object (a
black hole or neutron star); emission from the magnetospheric plasma of a
neutron stars (NS). The high brightness temperatures $T_{\rm b}\gtrsim
10^{35}\,$K of FRBs mean that the emission mechanism must be
coherent.
\citet{2018MNRAS.477.2470L} showed that models in the first category,
i.e. an outflow undergoing internal dissipation or interacting with
the surrounding medium, cannot reach typical FRB brightness
temperatures before the waves lose energy by induced Compton
scattering. On the other hand, if FRBs are produced within the
magnetosphere of NSs, the emission process is most likely powered by
the dissipation of magnetic energy, instead of rotational energy
\citep{2017ApJ...838L..13L, 2017ApJ...841...14M}.

The energy density of the FRB electromagnetic (EM) waves at radius $r$
from the source (in the limit $r\gg$ source size) is $U_{\rm
  EM} = L/(4\pi r^2 c)$, where $L$ is the isotropic equivalent
luminosity, and $c$ is the speed of light.
The magnetospheric B-field configuration at radius 
$r\gg R_*$ ($R_*\approx 10\,$km 
being the NS radius) is largely dipolar $B(r) \simeq
B_*(r/R_*)^{-3}$, where $B_*$ is the surface dipolar field. We require
the energy density  of the B-field $B^2/8\pi$ to be higher than
$U_{\rm EM}$ and obtain an upper limit for the radius of emission
\begin{equation}
  \label{eq:1}
  r\lesssim (6.2\times10^{7}\mr{\,cm})\, B_{*,15}^{1/2} L_{\rm
     45}^{-1/4},
\end{equation}
where $B_{*,15} = B_*/10^{15}\,$G and we use the highest inferred
isotropic equivalent luminosity of $L = 10^{45}L_{45}\rm\,erg\,s^{-1}$
as a fiducial value \citep{2017ApJ...841L..12B,
  2018MNRAS.475.1427B}. If the EM waves are powered only by   
particles' kinetic energy, the number density needs to be extremely high
$n\gtrsim (3\times10^{25}\mr{\,cm^{-3}})\,L_{45}r_7^{-2}
\gamma^{-1}(m/\me)^{-1}$, where $r_7 = r/10^7\,$cm, $\gamma$ is the mean 
Lorentz factor, and $m/\me$ is the rest mass of the particles divided by
electron mass. For any reasonable Lorentz factor, this number density
would make the source plasma extremely optically thick due to
free-free and/or curvature absorption \citep{2017MNRAS.468.2726K,
  2017arXiv170807507G} and radio waves cannot
escape. To circumvent this problem, we assume that the FRB waves are
emitted by the coherent curvature process when particles are
continuously accelerated by a quasi-static E-field parallel to the
local magnetospheric B-field, following \citet{2017MNRAS.468.2726K}.

In this \textit{Letter}, we show that FRBs should have a maximum
luminosity $L_{\rm max}$ because this parallel 
E-field must not exceed $\sim$5\% of the quantum critical field $E_{\rm
  c} = \me^2 c^3/(e\hbar)\simeq 4.4\times10^{13}\,$esu, where $\me$ 
and $e$ are the electron mass and charge, and $\hbar$ is the reduced
Planck's constant. Since the strength of the parallel E-field depends
on the location of the source plasma in the magnetosphere, we can use
$L_{\rm max}$ to constrain the source properties. In \S2, we derive
the upper limit of the parallel E-field and then 
calculate the maximum luminosity of FRBs. In \S3, we discuss the
effects of the maximum luminosity on the observed flux distributions
for repeating bursts 
from the same object and for the entire population of FRBs. In \S4,
we discuss some caveats of our simplified picture. Our main conclusions
are summarized in \S5. We use CGS units and the convention
$Q = 10^nQ_n$. All luminosities are in the
isotropic equivalent sense, unless otherwise explicitly stated. We use
the {\it Planck} best-fit cosmology \citep{2016A&A...594A..13P}.

\section{Luminosity Upper Limit due to Schwinger Pair Production}
We consider the situation of a quasi-static and uniform E-field and B-field near
the surface of a strongly magnetized NS, with $B\gg E$ and
$\b{E}\cdot\b{B}/B\ll E_{\rm c}$. It is possible to find an inertial frame where
the E-field is parallel to the B-field by applying a non-relativistic
Lorentz transform (in the original $\b{E}\times\b{B}$ direction). In this new
frame, the B-field strength is nearly unchanged and the E-field strength is
given by $E_{\para}\simeq \b{E}\cdot\b{B}/B$. It is well known that, when
$E_{\para}/E_{\rm c}\gtrsim 5\%$, the E-field will get quickly
shielded by copious Schwinger pairs and most of the energy
in the E-field gets converted into kinetic/rest-mass energy of
pairs \citep{Sauter1931, HeisenbergEuler36, Schwinger51}. For
completeness reason, we first re-derive the limiting E-field strength
\citep[following][]{2015arXiv150506400S} and then discuss the
implications on the maximum FRB luminosity.

The volumetric rate of pair production is given by
\citep[e.g.][]{2006PhRvD..73f5020K, 2010PhR...487....1R}
\begin{equation}
  \label{eq:4}
  \Gamma \simeq \alpha B E_{\para}/(\pi \hbar) \coth(\pi
  B/E_{\para})\mr{exp}(-\pi E_{\rm c}/E_{\para}),
\end{equation}
where $\alpha \simeq 1/137$ is the fine structure constant and
$\coth(\pi B/E_{\para})\simeq 1$ when $E_{\para}\ll \pi B$. Since
$\partial^2 E_{\para}/\partial t^2 = -4\pi \partial J/\partial t \simeq -8\pi
ec\Gamma$ (where $J$ is the current density), the timescale over which the
E-field is shielded is given by $\Delta t \simeq (\hbar/8ec\alpha B)^{1/2}
\mr{exp}(\pi E_{\rm c}/2E_{\para})$. When $E_{\para}\ll E_{\rm c}$,
this timescale is an extremely sensitive function
of $E_{\para}$, and the limiting E-field is
\begin{equation}
  \label{eq:3}
  E_{\para,\rm lim} \simeq {\pi E_{\rm c}\over \mr{ln}(8ec\alpha
    B\Delta t^2/\hbar)} \simeq {2.5\times10^{12}\mr{\,esu}
\over 1 + 0.018\mr{ln}(\Delta t_{-9}^2 B_{15})},
\end{equation}
where $\Delta t_{-9} = \Delta t/1\,$ns and $B_{15} =
B/10^{15}\,$G. We can see that the parallel E-field is quickly
shielded on sub-ns timescale when the parallel E-field exceeds
$2.5\times10^{12}\rm\,esu$.

In the following, we use simple arguments based on energy conservation
and source coherence to show that the strength of the parallel E-field
is directly related to the FRB luminosity. To generate waves of 
frequency $\nu$, the maximum source length in the longitudinal
direction is $\sim$$\lambdabar \equiv c/(2\pi \nu)$ in the NS
  rest-frame. Consider a source of longitudinal size $\lambdabar$ and
transverse size $\ell_\perp$, and moving along the local
magnetospheric B-field towards the observer at a Lorentz factor
$\gamma$ in the NS rest-frame. The 
local curvature radius of the B-field line is denoted as
$\rho$. For a fixed line of sight, the radiation formation length in
the NS rest-frame is $\rho/\gamma$, which corresponds to radiation
formation time of $\rho/(\gamma^2c)$ in the comoving frame of the
source. During this time, the EM fields or the influence by one
  particle on another travels a distance of $\rho/\gamma^2$ in the
  comoving frame, so the transverse size of the source (which is the
  same in the comoving frame and NS rest-frame)  is limited by
\begin{equation}
  \label{eq:9}
  \ell_\perp\lesssim \rho/\gamma^2.
\end{equation}
The emitting power of the source in the NS rest-frame is a
factor of $\sim$$\gamma^{-4}$ smaller\footnote{A factor of
  $\gamma^{-2}$ comes from relativistic beaming, and another factor
  of $\gamma^{-2}$ is because the difference between the speeds of
  photons and emitting particles is $\sim c/\gamma^2$ in the limit
  $\gamma\gg 1$.} than the
isotropic equivalent luminosity $L$ seen by the observer.
This emitting power is
supplied by $N\sim n\lambdabar\ell_\perp^2$ particles in the
coherent volume, where $n$ is the number density of radiating
particles in the NS rest-frame. From energy 
conservation, the emitting power of each particle in the NS rest-frame
is given by $E_\para e c$. Thus, we obtain
\begin{equation}
  \label{eq:10}
  \gamma^{-4}L \sim n\lambdabar\ell_\perp^2 E_\para e c,\
  \mr{or}\ L\sim (ne\lambdabar) (\ell_\perp\gamma^2)^{2} E_\para c.
\end{equation}
Since all \textit{radiating} particles in the coherent volume are of the same
charge sign (we ignore other background particles that do not
contribute to the observed FRB waves), we require that their 
Coulomb field does not exceed and shield the parallel E-field
--- the source of energy, i.e.,
\begin{equation}
  \label{eq:11}
  ne\lambdabar\lesssim E_\para.
\end{equation}
We insert inequalities (\ref{eq:9}) and (\ref{eq:11}) into
eq. (\ref{eq:10}) and obtain
\begin{equation}
  \label{eq:12}
  L\lesssim E_\para^2\rho^2 c.
\end{equation}
Using the upper limit of the parallel E-field $E_{\rm lim}$, we obtain
the maximum isotropic equivalent luminosity of an FRB
\begin{equation}
  \label{eq:5}
  L< L_{\rm  max,1}\sim
  (2\times10^{47}\mr{\,erg\,s^{-1}})\, \rho_6^2.
\end{equation}
We note that above maximum luminosity has no dependence on the
  Lorentz factor of the emitting particles. Below, we show that there
  is another Lorentz-factor-dependent maximum
  luminosity.

We assume that the emitting particles move close to the speed of light
($\gamma\gg 1$) along the magnetospheric B-field, and hence there is a
current density $nec$ parallel to the B-field. This current induces a 
transverse magnetic field $B_{\rm ind}\sim ne \ell_\perp$, which
must not perturb (or twist) the 
original B-field by more than a fraction of $\gamma^{-1}$ (the beaming
angle):
\begin{equation}
  \label{eq:13}
  ne \ell_\perp\lesssim B/\gamma.
\end{equation}
We insert the above inequality into eq. (\ref{eq:10}) and obtain
\begin{equation}
  \label{eq:14}
L\lesssim BE_\para \gamma\lambdabar (\ell_\perp\gamma^2)c\lesssim
BE_\para\gamma\lambdabar\rho c,
\end{equation}
where eq. (\ref{eq:9}) has been used in the second step. In the
coherent curvature emission model, the
radiation formation length is $\rho/\gamma\simeq \gamma^2\lambdabar$
in the NS rest-frame, so we obtain the typical Lorentz factor of
emitting particles to be $\gamma\simeq (\rho/\lambdabar)^{1/3}$. We
plug this Lorentz factor into eq. (\ref{eq:14}) and make use of
$E_\para<E_{\rm \para,lim} = 2.5\times10^{12}\,$esu, and 
then obtain
\begin{equation}
  \label{eq:15}
  L\lesssim L_{\rm max,2} \sim (2\times10^{46}\mr{\,erg\,s^{-1}})\,
  B_{15} \rho_6^{4/3}\nu_9^{-2/3},
\end{equation}
where $B_{15} = B/10^{15}\,$G is the B-field strength in
the source region and $\nu_9 = \nu/$GHz. The
strongest B-fields of NSs are believed to be produced due to
amplification by the $\alpha${--}$\Omega$ dynamo and may reach
$\mr{a\ few}\times 10^{17}\,$G, limited by the energy budget
of the differential rotation \citep{1993ApJ...408..194T}. Thus, the
inequality (\ref{eq:15}) may be weaker than
(\ref{eq:5}). Nevertheless, we combine these two conditions 
and obtain
\begin{equation}
  \label{eq:16}
  L\lesssim L_{\rm max} \sim (2\times10^{47}\mr{\,erg\,s^{-1}})\,
  \mr{min}(\rho_6^2, B_{16}\rho_6^{4/3}\nu_9^{-2/3}).
\end{equation}


\section{Observations}
In this section, we discuss the effects of $L_{\rm max}$ on the
observed flux distributions for repeating bursts
from the same object and for the entire population of FRBs.
\subsection{Repeating Bursts from the Same Object}
The luminosity function for the repeater FRB121102 is a power-law with
$\beta=1.7^{+0.3}_{-0.5}$ \citep{2017ApJ...850...76L} extending from
$\sim$$10^{40}$ to $\sim$$10^{43}\rm\,erg\,s^{-1}$, with (so-far) the brightest
one having peak flux $S=24\pm7\,$Jy \citep{2017ATel10693....1O}. One 
possible scenario is that the bursts are produced near the surface of
a NS where $\rho\sim 10\,$km and that the B-field strength near the
source is $\sim$$10^{15}\,$G (typical dipole surface field strength
inferred from Galactic magnetars). In this case, we have $L_{\rm
  max}\sim 10^{46}\rm\,erg\,s^{-1}$ and the observed 
flux distribution should have a cut-off at $\sim$$10^4\,$Jy. Note
that, if the bursts are produced far-away from the NS surface (but
within the light cylinder), then the B-field strength in the source
region is much weaker than that near the surface, and hence the cut-off
should show up at a lower flux level $\ll 10^{4}\,$Jy. In the
future, we may detect more 
repeaters, and then an interesting possibility is
that each repeating source may have different B-field strength and
curvature radius and hence different $L_{\rm max}$. We can see that the cut-off
luminosity $L_{\rm max}$ provides a powerful probe of the emission
location and the B-field strength near the source.

\subsection{The Entire FRB Population}
To show the observational effects of $L_{\rm max}$ for
the entire FRB population, we assume a global power-law luminosity
function in the form 
\begin{equation}
  \label{eq:6}
  {\d \dot{N}\over \d L} = \Phi(z) (\beta -1) L_{\rm  0}^{\beta-1}
L^{-\beta},
\end{equation}
where $L_{\rm  0}$ is a (fixed) reference luminosity, $\beta$ is
the power-law index, and $\Phi(z)$ is the normalization including the
cosmological evolution. We do
not assume that all FRBs repeat the same way as FRB121102, 
so the global power-law index may not be the same as the
repeater. Effectively, we treat each
repetition as a separate FRB originated from the same 
redshift. We leave $\beta$ as a free parameter 
between 1 and 2.5. The lower and upper limits are motivated by the observations
that brighter bursts are rarer than 
dimmer ones and that the DM distribution of known FRBs is not
concentrated near the lowest end. For simplicity, we also assume that
FRBs (on average) have a flat spectrum near $\sim$GHz frequencies
\citep{2018arXiv180404101G}, otherwise a receiver operating at a certain
frequency band will observe different parts of the intrinsic
spectrum for sources at different redshifts. This complication can be
effectively included in the $\Phi(z)$ factor and does not
significantly affect our calculations below.

In the ideal case of no propagation
effects such as scattering broadening, plasma lensing, absorption and
gravitational lensing (these complications will be discussed in \S 4),
the flux distribution of the observed bursts is 
\begin{equation}
  \label{eq:7}
  \begin{split}
    \dot{N}_{\rm det}(>S) =& \int_0^{z_{\rm max}} {\d z\over 1+z} {\d V
    \over \d z} 
\int_{4\pi D_{\rm L}^2 S}^{L_{\rm  max}}  {\d \dot{N}\over \d L} \d L \\
= &\int_0^{z_{\rm max}} {\d z\over 1+z} {\d V
    \over \d z} \Phi(z)\, \mr{max}\biggl[0, \\
&\left({4\pi D_{\rm L}^2 S\over L_{\rm  0}}\right)^{1-\beta} -
\left({L_{\rm  max}\over L_{\rm  0} }\right)^{1-\beta}   \biggr],
  \end{split}
\end{equation}
where $z_{\rm max}$ is the maximum redshift at which FRBs can be
produced, $\d V/\d z$ is the differential comoving volume within the
field of view for a certain telescope, $D_{\rm L}(z)$ is the
luminosity distance for redshift $z$, and $L_{\rm  max}$ is the
maximum isotropic equivalent luminosity of FRBs as given by
eq. (\ref{eq:16}). In the limit $L_{\rm  max}\rightarrow \infty$, the
$(L_{\rm max}/L_0)^{1-\beta}$ term in
eq. (\ref{eq:7}) vanishes, so we obtain a power-law
flux distribution $\dot{N}_{\rm det}(>S)\propto S^{1-\beta}$,
independent of the cosmological evolution of FRB
rate $\Phi(z)$. This is because the redshift distribution of
bursts in each flux bin $[S, S+\d S]$ is independent of
$S$. The flux distribution of the observed FRBs from Parkes
telescope is consistent with a single power-law but the
power-law index is not well constrained \citep[see the discussions
by][]{2016ApJ...830...75V, 2018MNRAS.475.1427B, 2018MNRAS.474.1900M},
due to the lack of a homogeneous sample with sufficient number of
bursts. For the luminosity function in eq. (\ref{eq:6}),
since $\beta > 1$, bursts near the cut-off luminosity are very 
rare and the only way to increase their detection rate is to use
telescopes with larger field of views.

\begin{figure}
  \centering
\includegraphics[width = 0.48 \textwidth,
  height=0.23\textheight]{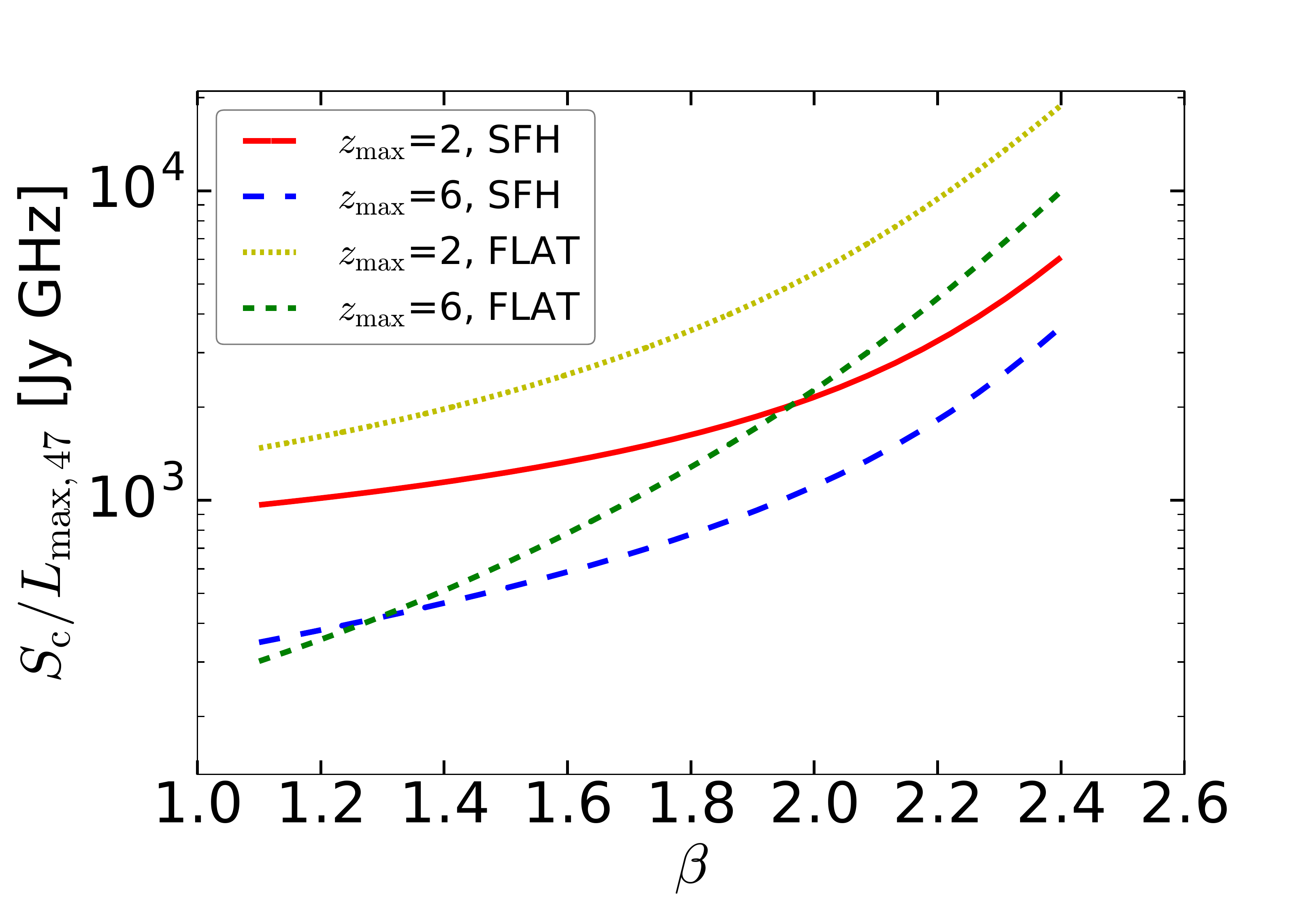}
\caption{The critical flux $S_{\rm c}$ (eq. \ref{eq:8}) as a function of
  the power-law index of the global luminosity function, for four
  different cases. For the red (solid) and blue (long-dashed) curves,
  we assume that the FRB rate tracks the cosmic star-formation history
  (SFH) given by \citet{2014ARA&A..52..415M}. For the 
  yellow (dotted) and green (short-dashed) curves, we assume a
  non-evolving (FLAT) FRB rate history. Two choices of maximum
  redshifts are shown $z_{\rm max} = 2$ and 6.
}\label{fig:Fcutoff}
\end{figure}

The critical flux at which the two terms on the RHS of
eq. (\ref{eq:7}) equal is given by
\begin{equation}
  \label{eq:8}
S_{\rm c} = 
\left[ {\int_0^{z_{\rm max}} {\d z\over 1+z} {\d V
    \over \d z} \Phi(z) ({4\pi D_{\rm L}^2})^{1-\beta}
  \over
\int_0^{z_{\rm max}} {\d z\over 1+z} {\d V
    \over \d z} \Phi(z)
}\right]^{{1\over\beta-1}} L_{\rm  max},
\end{equation}
which is linearly proportional to $L_{\rm  max}$ and depends on the 
power-law index
$\beta$, the cosmological rate evolution $\Phi(z)$, and the
maximum redshift $z_{\rm max}$. At flux levels much below $S_{\rm
  c}$, the flux source count is a power-law $\dot{N}_{\rm det}(>S)
\propto S^{1-\beta}$, but above this flux level, the deficit
of FRBs with $L\gtrsim L_{\rm  max}$ will be seen as a break in
the observed flux distribution. From
eq. (\ref{eq:7}), one can show that the distribution at $S\gg S_{\rm
  c}$ approaches $N(>S)\propto S^{-1.5}$ (Euclidean), since
bursts with $L\sim L_{\rm max}$ from the nearby Universe will dominate. In 
Fig. \ref{fig:Fcutoff}, we show the critical flux level
$S_{\rm c}$ as a function of the power-law index for four
different cases: (i) FRB rate $\Phi(z)$ either 
tracks the cosmic star-formation history or is non-evolving
throughout the history; (ii) the maximum redshift $z_{\rm max}$ is either 2
or 6. The choices of $z_{\rm max}$ is motivated by the highest redshift of
$z\sim 2$ inferred from the DM of FRB 160102
\citep{2018MNRAS.475.1427B}. We find that 
the value of $S_{\rm c}$ has a weak dependence on the cosmic
evolution of FRB rate and that the dependence on the
power-law index $\beta$ is also fairly mild (varying by about one order of
magnitude).

Therefore, for a power-law global luminosity function, we predict
the cumulative flux distribution to be $\dot{N}_{\rm det}(>S)\propto
S^{1-\beta}$ below the flux level $S_{\rm c}\sim(10^3${--}$10^4)L_{\rm
   max,47}\,$Jy and $\dot{N}_{\rm det}(>S)\propto
S^{-1.5}$ at $S\gg S_{\rm c}$. The deficit of high flux FRBs should be
noticeable with sufficiently large number of detections near and above $S\sim
S_{\rm c}$. The cut-off luminosity $L_{\rm max}$ can be inferred
from the critical flux $S_{\rm c}$ via eq. (\ref{eq:8}).

Unfortunately, the expected all-sky rate of FRBs near $S_{\rm
  c}$ is highly uncertain, mainly because the power-law index $\beta$
is only weakly constrained by current data. From the Parkes FRB
sample, \citet{2018MNRAS.475.1427B} inferred a rate\footnote{We note
  that the reported fluxes in their sample are  
based on the assumption that the bursts occurred within the half-power
width of the discovery beam. It was later realized that,
  at discovery, FRB121102 (the repeater) was in a side
   lobe where the sensitivity is $\sim$10\% of that at the beam center
   \citep{2014ApJ...790..101S, 2017Natur.541...58C}. Thus, the
   locations of some Parkes bursts may also be in the side lobes and 
hence their true fluxes are higher than those reported.  Since the
effective field of view (including side lobes) is larger, 
 this will give a lower all-sky rate $\dot{N}_{\rm det}(\gtrsim S_{\rm
   th, eff})$ above a higher effective completeness threshold flux $S_{\rm
   th,eff}$. 
} of $\dot{N}_{\rm det}(\gtrsim S_{\rm th}) \sim 2\times10^{3}
\rm\,sky^{-1}\,d^{-1}$ above the completeness threshold flux $S_{\rm th}\sim
3\mr{\,Jy\,GHz}$. Taking their rate at face value, we expect the
all-sky rate near the flux level $S_{\rm
  c}\sim3\times10^3L_{\rm max,47}\rm\,Jy\,GHz$ to be $\dot{N}_{\rm exp}(\gtrsim S_{\rm
  c})\sim 2\times10^{3(2-\beta)}L_{\rm
  max,47}^{1-\beta}\rm\,sky^{-1}\,d^{-1}$. For $L_{\rm max} =
10^{47}\rm\,erg\,s^{-1}$ and 
$\beta = 1.7$ (or 2.3), the product of solid angle and observing time
per FRB detection with $S\sim S_{\rm c}$ is $\sim$$20\rm\,sr\cdot hr$
(or $1.2\times10^3\rm\, sr\cdot hr$).



\section{Discussion}

In this section, we discuss some caveats of our simplified picture. As
more data accumulates, they may become important issues
to look at in detail in future works.

(1) The signal-to-noise ratio of an FRB is determined by a combination of
flux $S$ and duration $\tau$ as $\mr{SNR}\propto S\sqrt{\tau}$
\citep{2015MNRAS.447.2852K}. In eq. (\ref{eq:7}), the luminosity at
the detection threshold $4\pi
D_{\rm L}^2 S$ (for a given redshift and 
flux) should include an additional factor $\propto \tau^{-1/2}$ and
then we integrate over the intrinsic distribution of burst
durations. We can see that the shape of the flux distribution function
$\d \dot{N}_{\rm det}/\d S$ stays the same, as long as the intrinsic
distribution of burst durations is not correlated with their
luminosities (such a correlation has not been found in the 
literature).

(2) The observed flux $S_{\rm obs}$ may be different
from the intrinsic/unattenuated flux $S = L/4\pi D_{\rm L}^2$ for a given
redshift and luminosity. When there is significant scattering broadening,
intra-channel dispersion smearing, absorption, insufficient time
resolution, or when the location of the burst is far away from the 
center of the discovery beam, we have $S_{\rm obs}<S$. On the other
hand, magnification bias due to lensing of FRBs by plasma structures
in the  host galaxies \citep{2017ApJ...842...35C} may lead to $S_{\rm
  obs}>S$ for a fraction of the observed bursts. Thus,
the critical flux above which the luminosity function cut-off is noticeable in the
source count may be different than the unattenuated flux
$S_{\rm c}$ in eq. (\ref{eq:8}). These effects make it harder to infer
the maximum luminosity $L_{\rm max}$ from observations, but the
existence of a cut-off in the luminosity function can still be tested.

(3) We have assumed the luminosity function to be a single power-law
with a cut-off at $L_{\rm  max}$ and the power-law index
to be in the range (1, 2.5). For other luminosity function models,
eqs. (\ref{eq:7}) and (\ref{eq:8}) are generally valid. For instance,
an alternative luminosity function is a broken power-law and in this
case $\beta\geq2.5$ is allowed on the high-luminosity end  (as is the
case of long gamma-ray bursts). Another possibility is that there is
another cut-off at the low luminosity end. In these cases, it is
straightforward to solve eq. (\ref{eq:8}) for the critical flux $S_{\rm c}$ (which
may be significantly different from that shown in
Fig. \ref{fig:Fcutoff}) and determine where the deficit of
high-luminosity FRBs above $L_{\rm max}$ will show up in the observed
flux distribution $\d\dot{N}_{\rm det}/\d S$.


(4) The observed flux distribution suffers from magnification bias due to
 strong gravitational lensing by intervening galaxies. For FRBs at $z\sim2$ (near the
peak of the cosmic star-formation history), the optical depth for
large magnification $\mu\gg 1$ is roughly $P(>\mu)\sim
10^{-3}\mu^{-2}$ \citep[e.g.][]{2011ApJ...742...15T}, which should be
multiplied by the luminosity function $\d\dot{N}/\d \mr{ln}L\propto
L^{1-\beta}$ to calculate the contribution to 
the source count at a given flux. If $\beta < 3$, then the
majority of the lensed sources with apparent luminosity $\gg L_{\rm
   max}$ come from those sources with intrinsic luminosity $L\sim
 L_{\rm  max}$ \citep{1992ARA&A..30..311B}. Thus, 
the observed flux distribution of lensed (L) FRBs should be 
$\dot{N}_{\rm det, L}(>S)\propto S^{-2}$ above the critical flux
$S_{\rm c}$, which is steeper than $\dot{N}_{\rm det, NL}(>S)\propto
S^{-1.5}$ for unlensed (NL) FRBs at $S\gg S_{\rm c}$. Therefore, the
unlensed population always dominate at all flux levels and
magnification bias should not a serious problem for constraining the
cut-off luminosity $L_{\rm  max}$. 

\section{Summary}

In this \textit{Letter}, we provide a novel way to test the model that
FRBs are from coherent curvature emission powered by the dissipation
of magnetic energy in the magnetosphere of NSs. In this model, the emitting particles 
are continuously accelerated by a quasi-static E-field parallel to
the local B-field. We use simple arguments based on
energy conservation and source coherence to show that the isotropic
equivalent luminosity of an FRB is directly related to the parallel
E-field strength. When this parallel E-field
exceeds about 5\% of the quantum critical field strength, it is quickly
shielded by Schwinger pairs on sub-ns timescales (and hence the FRB
emission cannot be sustained). Based on this limiting E-field, we
show that there is a maximum isotropic equivalent luminosity of $L_{\rm
  max}\sim (2\times 10^{47}\, \mr{erg\,s^{-1}})\,
\mr{min}(\rho_6^2, B_{16} \rho_6^{4/3}\nu_9^{-2/3})$, where $\rho$ is the
curvature radius of the magnetic field lines near the source
region. Future observations can measure $L_{\rm max}$ and hence probe
the source location and B-field strength. 

For the repeater FRB121102, this cut-off
luminosity corresponds to a maximum flux of $S_{\rm max} = L_{\rm
  max}/4\pi D_{\rm L}^2\sim 10^5L_{\rm max,47}\,$Jy. Each repeating
source may have a different $L_{\rm max}$ from the others, depending
on the source location and B-field strength. We encourage monitoring
the repeater for an extended amount of time with a low-sensitivity
telescope. 

If the entire population of FRBs has a global
luminosity function, then the cut-off luminosity $L_{\rm max}$ should
be observable as a deficit of high-flux FRBs in the the observed flux
distribution. Taking the simplest case of a 
power-law luminosity function $\d N/\d L\propto L^{-\beta}$ as an
example, we show that there is a 
critical flux $S_{\rm c}\sim (10^3${--}$10^4)L_{\rm
  max,47}\,$Jy, below and above which the cumulative flux distribution
will be $\dot{N}_{\rm det}(>S)\propto S^{1-\beta}$ (for $S\ll S_{\rm c}$)
and $\dot{N}_{\rm det}(>S)\propto S^{-1.5}$ (for $S\gg S_{\rm c}$). Bright
FRBs near or above the critical flux $S_{\rm c}$ have a much lower all-sky
rate than those currently detected. Extrapolating the rate of
  Jy-level FRBs to higher fluxes and assuming $L_{\rm max} =
  10^{47}\rm\,erg\,s^{-1}$, we estimate the 
detection rate of bright FRBs near $S_{\rm c}$ by ASKAP 
\citep[sky coverage $\Omega/4\pi\sim 4\times10^{-3}$ at 0.7-1.8 
GHz,][]{2017ApJ...841L..12B} to be $0.06\rm\,d^{-1}$ for $\beta = 1.7$ and
$0.001\rm\,d^{-1}$ for $\beta = 2.3$. The rate for CHIME \citep[sky coverage
$\Omega/4\pi\sim 7\times10^{-3}$ at 
400{--}800 MHz,][]{2018arXiv180311235T} may be slightly higher. We
encourage searching for ultra-bright FRBs by low-sensitivity
telescopes with large field of views.

\section{acknowledgments}
We thank Vikram Ravi for useful discussions. We also thank the referee
for comments which improved the clarity of the presentation.
This research benefited from interactions at the ZTF Theory Network
Meeting, funded by the Gordon and Betty Moore Foundation through Grant
GBMF5076. W.L. was supported by the David and Ellen Lee Fellowship at
Caltech.

\bibliographystyle{mnras}
\bibliography{refs}

\label{lastpage}
\end{document}